\documentclass{PoS}

\newcommand{\snn}          {\ensuremath{\sqrt{s_{\mathrm{NN}}}}}
\newcommand{\GeVc}         {GeV/$c$}
\newcommand{\pT}           {\ensuremath{p_{\mathrm{T}}}}

\title{Machine Learning based jet momentum reconstruction in Pb--Pb collisions measured with the ALICE detector}

\ShortTitle{ML-based jet momentum reconstruction in Pb--Pb}

\author{\speaker{R\"udiger Haake} for the ALICE Collaboration\\
        Yale University, Wright Laboratory, New Haven, CT, USA\\
        E-mail: \email{ruediger.haake@cern.ch}}

\abstract{
The precise reconstruction of jet transverse momenta in heavy-ion collisions is a challenging task. A major obstacle is the large number of uncorrelated (mainly) low-$p_\mathrm{T}$ particles overlaying the jets. Strong region-to-region fluctuations of this background complicate the jet measurement and lead to significant uncertainties.
We developed a novel approach to correct jet momenta (or energies) for the underlying background in heavy-ion collisions. The approach allows the measurement of jets down to extremely low transverse momenta and for large resolution $R$ by making use of common Machine Learning techniques to estimate the jet transverse momentum based on several parameters.

In this conference proceeding, we will present transverse momentum spectra and nuclear modification factors of track-based jets that have been corrected by this Machine Learning approach and comparisons to published results where possible. The analysis was performed on Pb--Pb collisions at $\sqrt{s_\mathrm{NN}} = 5.02$ TeV recorded with the ALICE detector and measures jets with large resolution parameters for low momenta, unprecedented thus far in data on heavy-ion collisions.
}

\FullConference{
European Physical Society Conference on High Energy Physics - EPS-HEP2019 -\\
			10-17 July, 2019\\
			Ghent, Belgium}

\begin{document}

%%%%%%%%%%%%%%%%%%%%%%%%%%%%%%%%%%%%%%%%%%%%%%%%%%%%%%%%%%%%%%%%%%%%%%%%%%%%%%%%
\section{Introduction}
% Background introduction
The reconstruction of particle jets in heavy-ion collisions is a complex task. The main obstacle is the overwhelmingly large background of particles that do not originate from hard interactions.
In ALICE~\cite{ALICE2008}, the mean momentum density in 0--10\% most central collisions at $\snn =2.76$~TeV leads to a contribution to the jet momentum that is already of the order of the typical jet momentum itself. The average charged particle transverse momentum density for particles with momenta above $0.15$ \GeVc\ is $\langle \rho \rangle \approx 138.2$ \GeVc\ per unit area, while its standard deviation is $\sigma(\rho) = 18.5$ \GeVc\ ~\cite{Background2012}. Since jets are rare objects, these numbers already provide a good estimate of the mean background in the selected events.
In addition, this background shows large uncorrelated and correlated region-to-region fluctuations. Uncorrelated fluctuations are due to random Poissonian fluctuations of the number of particles and their momenta. Sources of correlated fluctuations are e.g.\ physical correlations of particles from hydrodynamic flow or non-uniform detector acceptances.
These fluctuations have a large impact on the reconstructed jet momentum and on the jet axes by directly affecting the jet finding algorithm and eventually result in large uncertainties on the final measurements.
An approach to at least lower the impact of the background at the expense of a potential fragmentation bias is a higher $\pT$-cut for constituents used in the jet finding algorithm. This massively reduces the background, which mostly consists of low-$\pT$ particles, but it also discards the low-$\pT$ parts of the jet.
Note that while the exact numbers for the background properties given above differ for $\snn = 5.02$~TeV, -- due to the higher multiplicity, the numbers are slightly larger -- the overall picture and conclusions stay the same.

% Background method and motivation for new method
In the de-facto standard method for jet spectra measurements in ALICE, the background momentum density per unit area is calculated on an \textit{event-by-event} basis. Each jet is then corrected by the average momentum density in the event multiplied by the jet area.
The area-based method corrects the jet momentum for the average background but leads to large residual fluctuations.
These residual fluctuations are then typically corrected for on a statistical basis in an unfolding procedure, see for instance~\cite{ChargedJetRAA2014}.

The new approach used in this analysis was developed in~\cite{MLMethodPaper2019} and calculates the corrected jet momentum on a \textit{jet-by-jet} basis to reduce the residual fluctuations and to allow a more precise estimate for the jet momentum. This enables the measurement of jets in heavy-ion collisions with transverse momenta much lower than what is currently possible with the area-based method. Here, common Machine Learning~(ML) techniques are applied to obtain the mapping between jet parameters, e.g.\ constituent momenta, and the \textit{true} transverse momentum of the jet.

The following analysis strategy is proposed to train, validate, and apply the model to real data.
The first step is the creation of the toy model data.
The second step is the training of the ML-based estimators and their evaluation on the toy model.
Training and evaluation datasets are independent subsamples of the full toy model dataset.
During this step, the model hyperparameters were adjusted to obtain a good performance.
These first two steps are presented in the method paper cited above.
In a third step, the trained and commissioned estimator is applied to heavy-ion data to obtain background-corrected spectra, here jet momentum spectra.
In addition, a response matrix is created to unfold residual fluctuations (and possible detector effects). It can also account for potential biases in the method.
The response matrix is created by embedding vacuum jets into a data background.
The last step is the unfolding procedure to measure the final spectra.

%%%%%%%%%%%%%%%%%%%%%%%%%%%%%%%%%%%%%%%%%%%%%%%%%%%%%%%%%%%%%%%%%%%%%%%%%%%%%%%%
\section{Datasets and jet reconstruction}

The main result shown here is based on Pb--Pb data that was taken in 2015 at $\snn = 5.02$ TeV with a minimum bias trigger setup.
For 0-10\% and 30-50\% most central collisions, 7M and 14M events have been processed for further analysis, respectively.
For the response, a PYTHIA8~\cite{PYTHIA2006} simulation was used which was reconstructed to the detector level using GEANT3~\cite{GEANT1994}. 
The simulation has the same detector configurations as in the real Pb--Pb dataset.
For the reference of the nuclear modification factor and comparisons, similar observables measured in pp collisions were used. The datapoints were taken from~\cite{PP} without further modification.

The presented analysis is on track-based jets, which are clustered by a jet finding algorithm into jets. Since jet reconstruction can be sensitive to holes or lowered efficiencies in the acceptance, tracks from different reconstruction classes are combined such that they exhibit a uniform efficiency in the full acceptance. In addition, the tracks are required to fulfill $\eta < 0.9$ and $p_\mathrm{T} > 0.150 \mathrm{~GeV}/c$.
The FastJet package~\cite{FastJet2006} was used to clusterize tracks with the anti-$k_\mathrm{T}$ and $k_\mathrm{T}$ algorithms~\cite{AntiKT2008}, to reconstruct signal jets and for use in the area-based background estimate in the latter case. $R=0.2,\;0.4$, and $0.6$ were used as jet resolution parameters. Further cuts were applied to jets before analysis:
First, to assure that jets are contained within the acceptance to avoid edge effects, $\eta_\mathrm{jet} < 0.9 - R$ was demanded.
Second, jets containing tracks with $p_\mathrm{T} \geq 100$~\GeVc\ were discarded.
Last, only jets with $A_\mathrm{jet} > 0.557\pi R^2$ were accepted.
%In the presented method, the jet definition -- i.e.\ its position and constituents -- is not changed.

%%%%%%%%%%%%%%%%%%%%%%%%%%%%%%%%%%%%%%%%%%%%%%%%%%%%%%%%%%%%%%%%%%%%%%%%%%%%%%%%
\section{Background estimator}

The core of this analysis is the correction of the jet $\pT$ with a Machine-Learning-based regression algorithm based on several low- and high-level input parameters.
In essence, a regression model is trained to perform the mapping $p_\mathrm{T, raw} \rightarrow p_\mathrm{T, rec}$
and is applied on a jet-by-jet basis. A supervised learning approach is used, i.e.\ samples with known truth are given as examples to train the model.
The training data is based on PYTHIA jets, reconstructed to detector-level, embedded into a thermal model.

Several Machine Learning techniques were tested to perform the mapping of raw and reconstructed momentum, including linear regression, random forests, and neural networks.
Training and evaluation was performed using the Python package \texttt{scikit-learn}~\cite{SKLearn}. The baseline model in this analysis is the neural network with three hidden layers with 100, 100, and 50 neurons, respectively, but all considered models perform better than the area-based background correction method.
In~\cite{MLMethodPaper2019}, the method is described in detail alongside with toy studies. For brevity, only a brief description of the method and the technical setup is given here.

\subsection{Training datasets}

To create events with particle jets in a heavy-ion background, PYTHIA-generated events are embedded in a thermal background.
The events are from the above mentioned PYTHIA production at $\sqrt{s} = 5.02$~TeV, reconstructed to detector-level.

The thermal background is created by randomly distributing charged particles according to a flat particle multiplicity distribution ranging from $0$ to $3000$ tracks and with a realistic (thermal) momentum distribution. A flat multiplicity distribution contains sparse and dense events with equal weights and the maximum track count of $3000$ is representative of the most central events.
The momentum distribution is defined such that it coincides with the real track momentum distribution at low $\pT$.
For higher particle momenta roughly above 4~\GeVc, the background momentum distribution falls much quicker, i.e.\ exponentially.

\subsection{Input parameters and regression target}

In order to find a suitable combination of input parameters, the analysis was repeated for a large variety of parameter sets. The number of parameters used is kept small to avoid a dependence on data subtleties. Eventually, the following input parameters prove to be useful, discriminative features:
the jet transverse momentum, corrected by the established area-based method,
jet angularity,
the number of constituents within the jet,
and the transverse momenta of the first eight leading, i.e.\ hardest, particles within the jet.

For the training, the supervised learning techniques that are applied need a truth value assigned to each sample, i.e.\ to each jet.
The truth that is approximated by the correction method -- the regression target -- is the true jet momentum on detector level.
Here, it is defined as the reconstructed jet momentum multiplied by the momentum fraction that is carried by PYTHIA particles in the jet.

\subsection{Performance}

\begin{figure}
\begin{center}
  \includegraphics[width=0.43\textwidth]{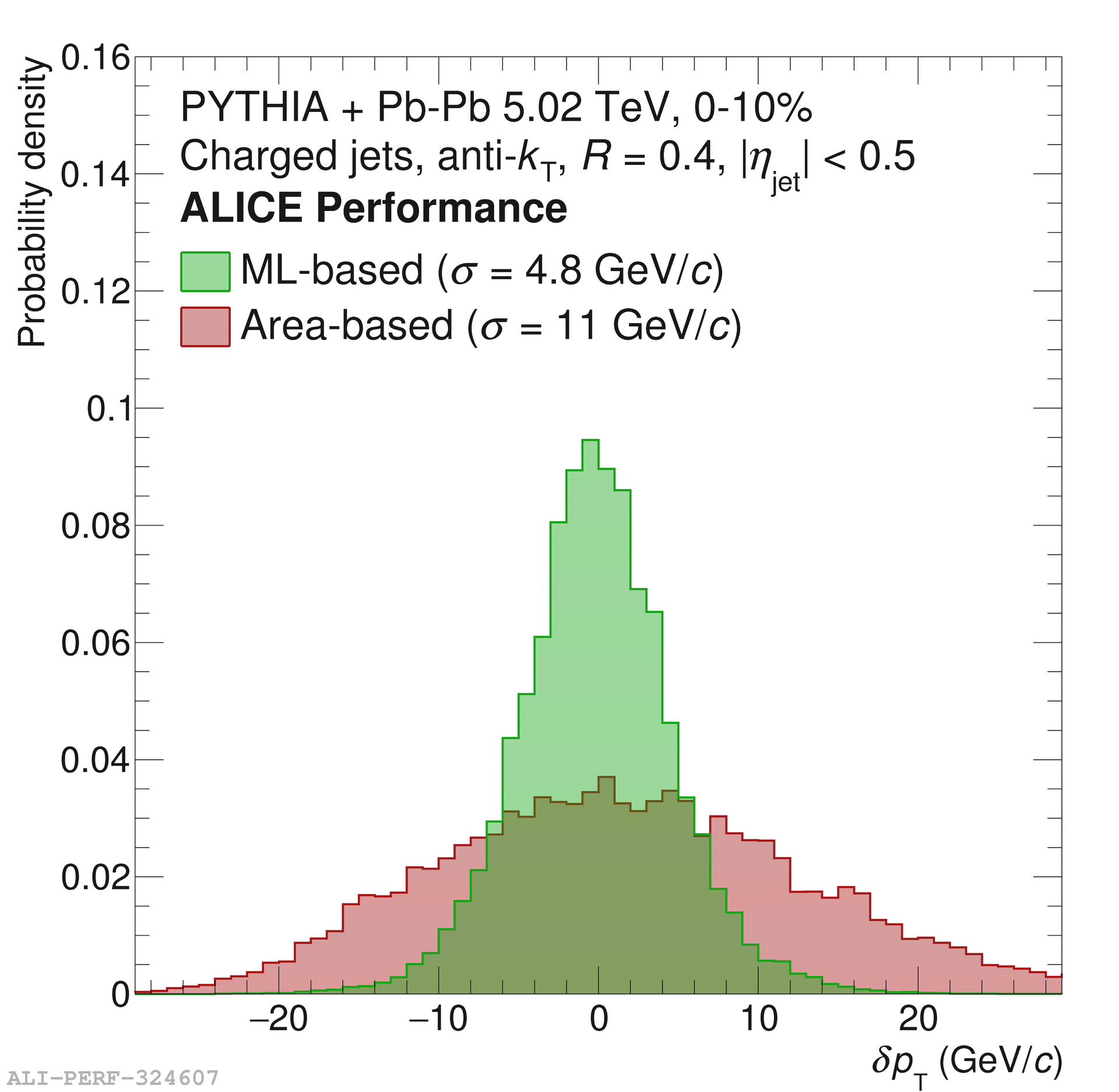}
  \includegraphics[width=0.43\textwidth]{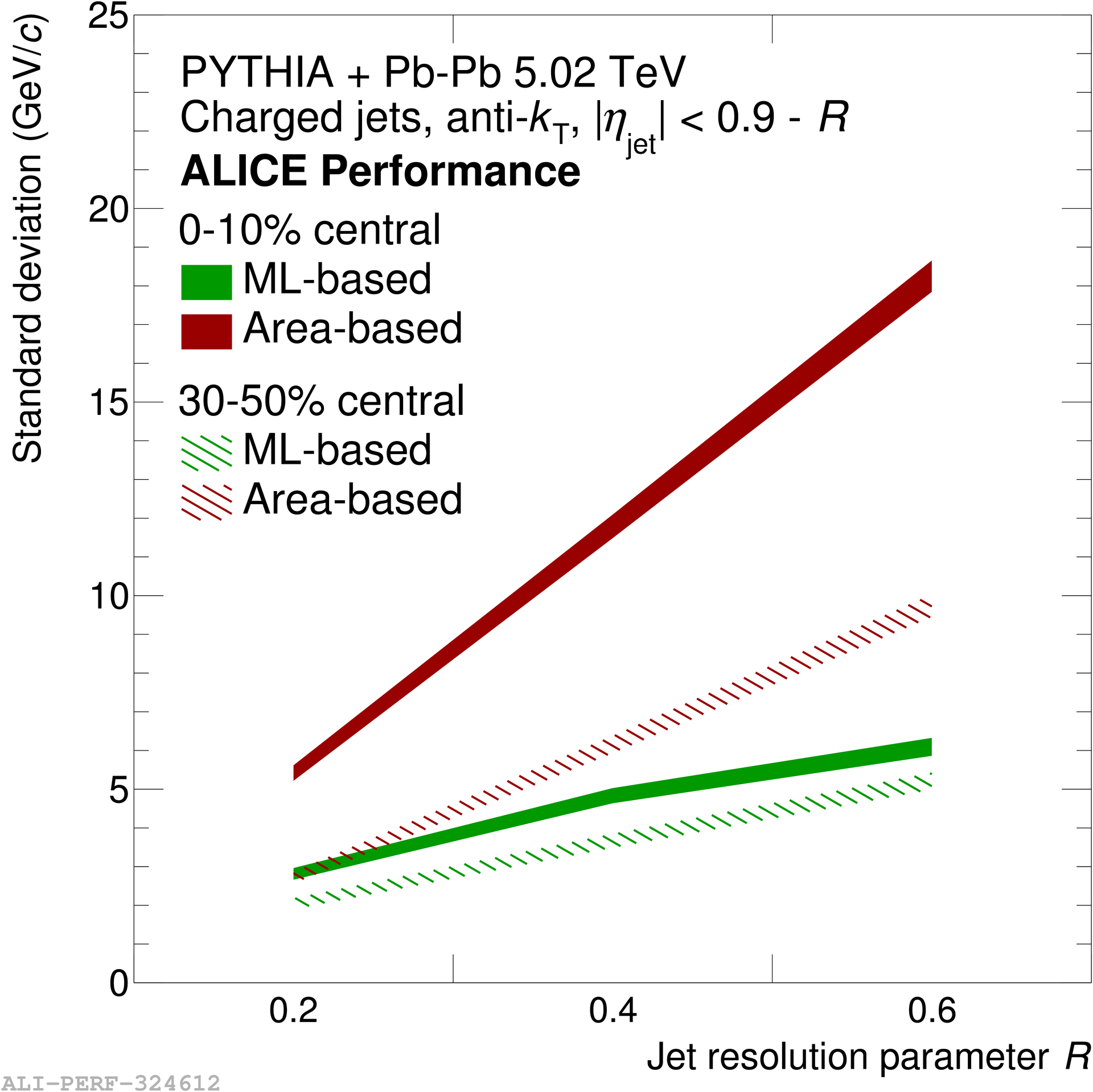}
  \caption{Residual $\pT$-distributions of embedded jet probes of known transverse momentum.}
  \label{fig:deltaPt}
\end{center}
\end{figure}

To compare the new background estimator to the area-based estimator, an embedding procedure is used: Jet probes with known transverse momentum are embedded into real Pb--Pb events. The residual difference between background-corrected $\pT$ and true probe $\pT$ is a direct measure for how precise the background is approximated.
The left plot in Fig.~\ref{fig:deltaPt} shows the comparison of the different background estimators for $R=0.4$.
In the right plot, the standard deviation of residual distributions for different $R$ for both background estimators and different centralities is shown as a measure for the width of this distribution. From these plots, a much better performance for the background estimator is expected for the new ML-based approach.

% Delta pT distributions

%%%%%%%%%%%%%%%%%%%%%%%%%%%%%%%%%%%%%%%%%%%%%%%%%%%%%%%%%%%%%%%%%%%%%%%%%%%%%%%%
\section{Results}

In this analysis, three observables are of central interest. The track-based jet production spectrum, nuclear modification factor $R_\mathrm{AA}$, and the jet cross-section ratio.
The nuclear modification factor is defined as the ratio of the per-event jet yields in Pb--Pb and the cross sections in pp multiplied by $T_\mathrm{AA}$ that accounts for the collisions geometry. It is
\begin{equation}
\label{eq:RAA}
R_\mathrm{AA} = \frac{1}{N_\mathrm{ev,\,PbPb} \cdot T_\mathrm{AA}} \frac{\mathrm{d}^{2}N_\mathrm{ch\;jet}/\mathrm{d}{p}_\mathrm{T, ch\;jet} \mathrm{d}{\eta}_\mathrm{jet}}{{\mathrm{d}^{2}\sigma_\mathrm{pp}/\mathrm{d}{p}_\mathrm{T, ch\;jet} \mathrm{d}{\eta}_\mathrm{jet}}},
\end{equation}
where $T_\mathrm{AA} = N_\mathrm{coll}/\sigma_\mathrm{tot}$~\cite{TAANote}.
It is a direct measure for the nuclear suppression of reconstructed jet yields in Pb--Pb compared to the pp reference.
The jet cross-section ratio $\sigma(R_1)/\sigma(R_2)$ is simply defined as the ratio of the differential jet production cross sections or, alternatively, as the per-event jet yields in Pb--Pb for different resolution parameters.

Systematic uncertainties have been calculated for each observable and setting separately by considering several variations. The uncertainties derived from these variations are assumed to be symmetric and independent of each other. Therefore, the full systematic uncertainty is a quadratic sum of the single uncertainties.
The following variations have been taken into account: Tracking efficiency uncertainty, unfolding method, regularization parameter, prior, measured $\pT$-range, fragmentation.
The uncertainty on the fragmentation takes into account the fact that the estimator was trained on PYTHIA jets. Since the jet fragmentation might differ in certain regions of the phase space in heavy-ion collisions, the impact of a different fragmentation on the final results is estimated. In this case, a different response matrix assuming quark-jet fragmentation is used in the unfolding procedure.

\begin{figure}
  \includegraphics[width=0.499\textwidth]{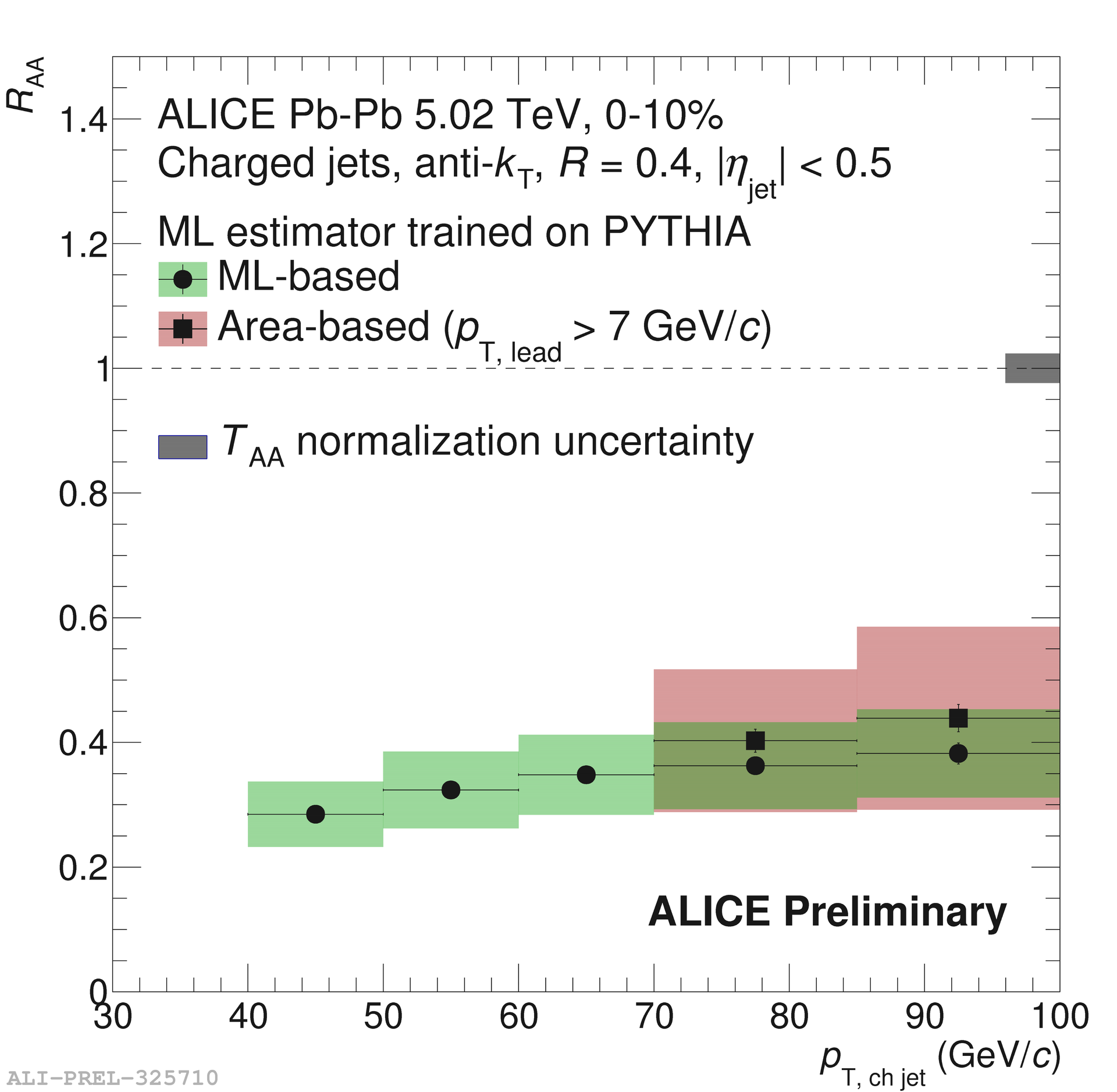}
  \includegraphics[width=0.499\textwidth]{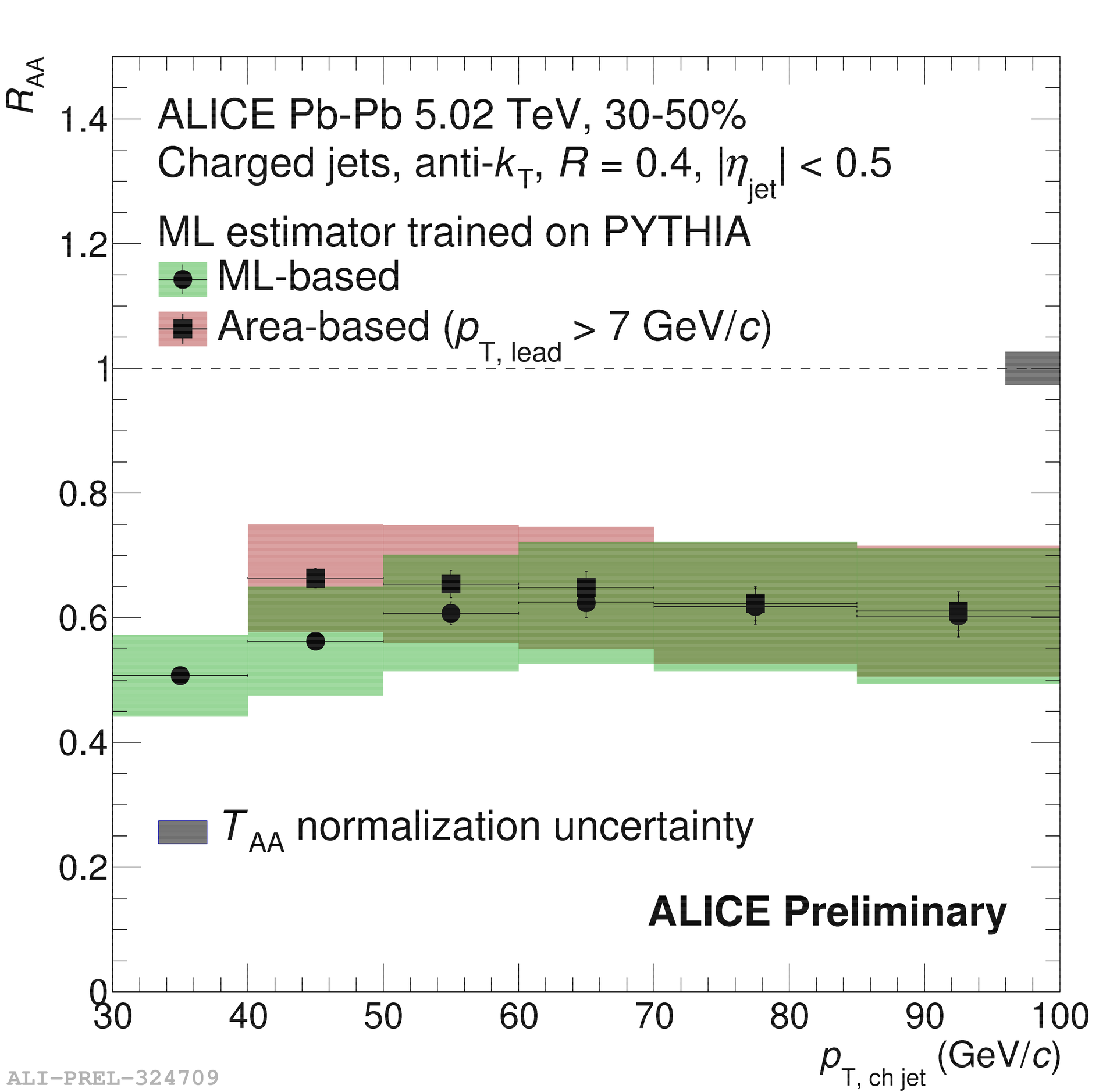}
  \caption{Nuclear modification factor for $R=0.4$, comparing spectra corrected with either ML-based or area-based estimator. 0-10\% (left) and 30-50\% (right) most central collisions.}
  \label{fig:RAA_1}
\end{figure}

\begin{figure}
  \includegraphics[width=0.499\textwidth]{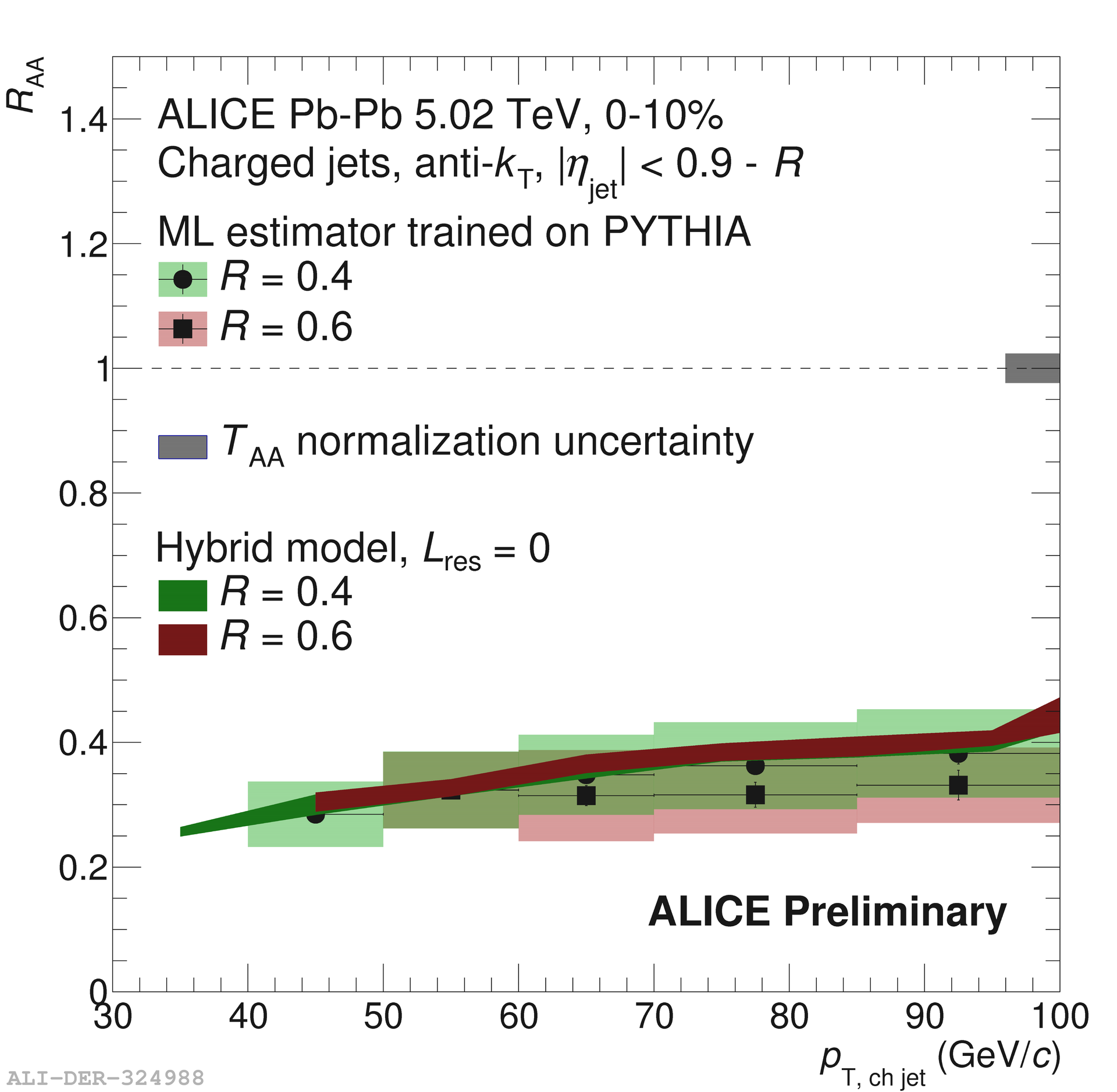}
  \includegraphics[width=0.499\textwidth]{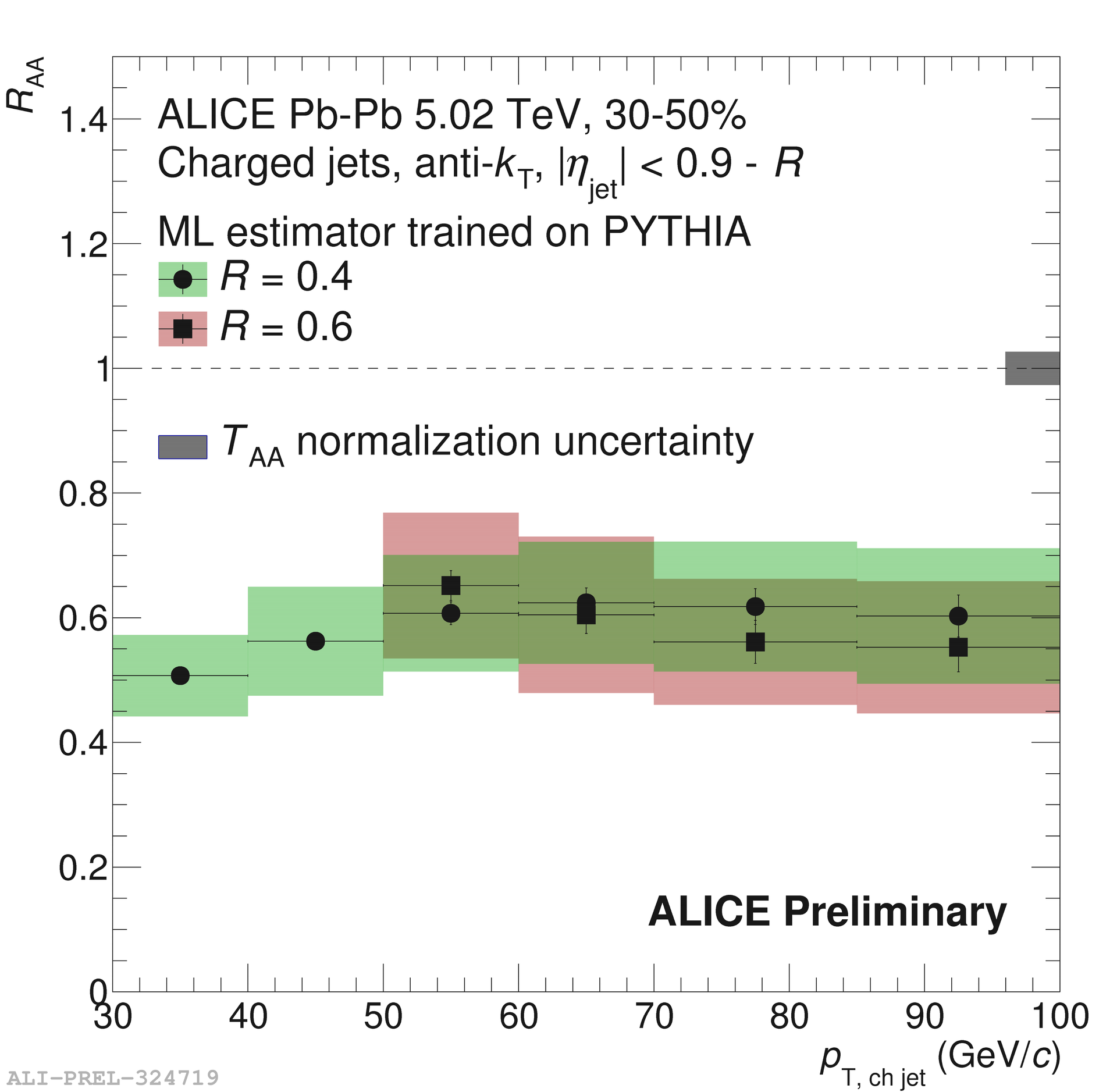}
  \caption{Nuclear modification factor for $R=0.4$ and $R=0.6$ for 0-10\% (left) and 30-50\% (right).}
  \label{fig:RAA_2}
\end{figure}

\begin{figure}
  \includegraphics[width=0.499\textwidth]{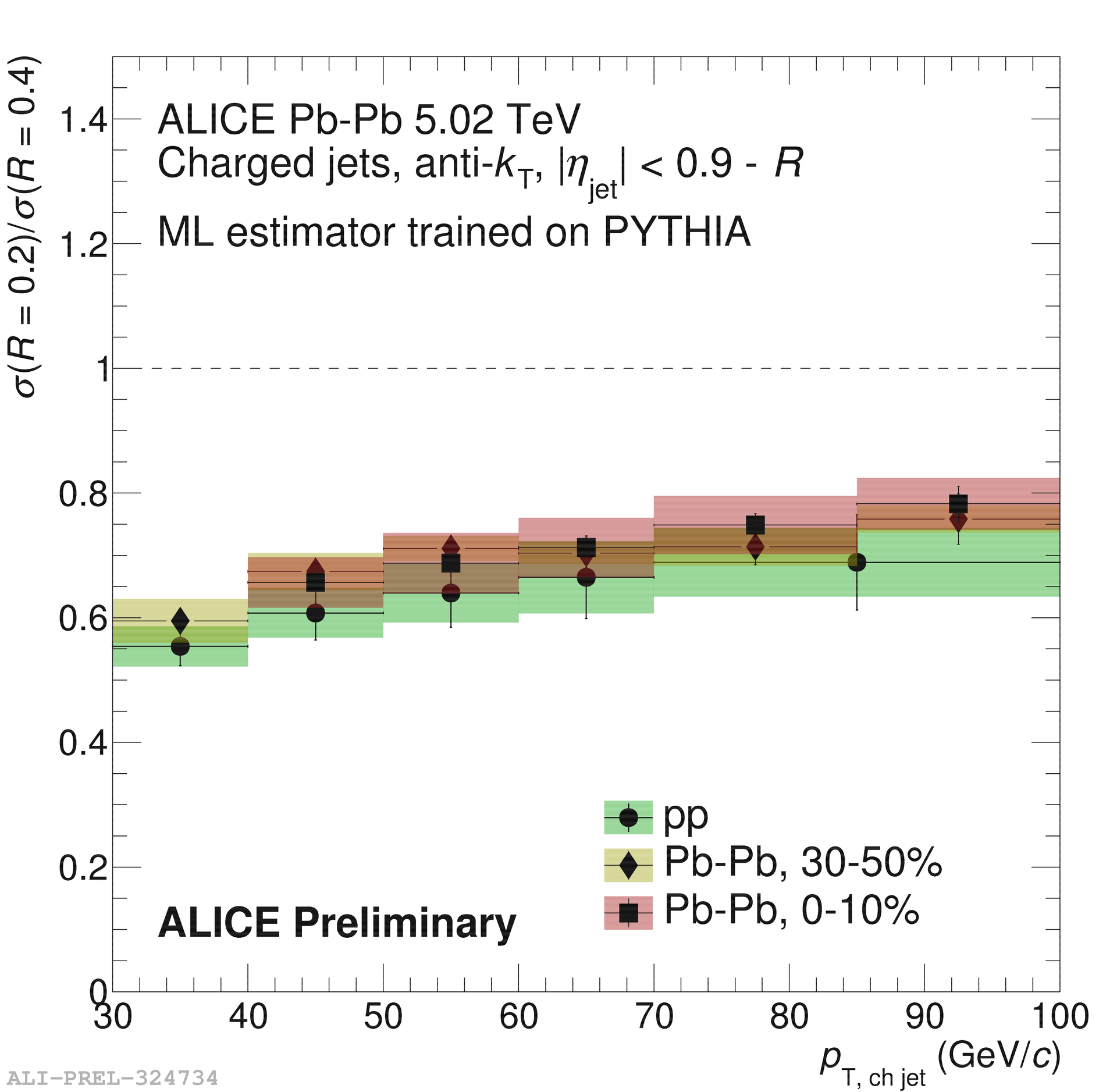}
  \includegraphics[width=0.499\textwidth]{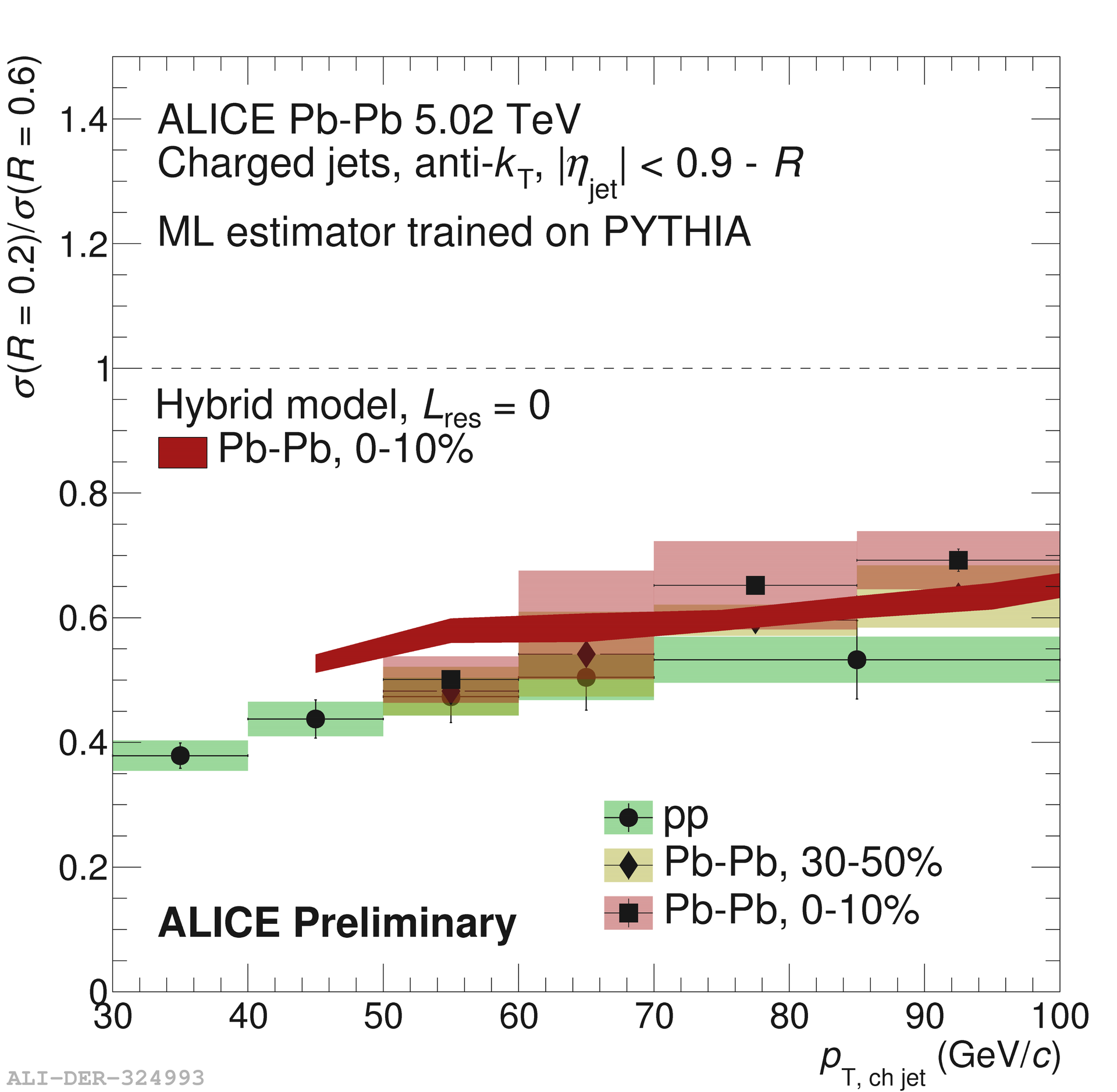}
  \caption{Jet cross-section ratios for $R=0.2/R=0.4$ (left) and $R=0.2/R=0.6$ (right).}
  \label{fig:CSR}
\end{figure}

The results are presented in Figs.~\ref{fig:RAA_1},~\ref{fig:RAA_2}, and~\ref{fig:CSR}.

% RAA, comparison old vs. new estimator

Figure~\ref{fig:RAA_1} proves that the new estimator yields fully compatible results for $R=0.4$, compared to the area-based estimator and for 0-10\% and 30-50\% most central collisions. Besides a higher precision, also a much larger range towards low $\pT$ can be achieved. In Fig.~\ref{fig:RAA_2}, the nuclear modification factor is shown for $R=0.4$ and, for the first time in heavy-ion collisions at the LHC, for $R=0.6$. No significant $R$-dependence can be observed and the data for 0-10\% most central collisions nicely coincides with Hybrid Model calculations~\cite{HybridModel}.

In Fig.~\ref{fig:CSR}, the jet cross-section ratio is shown for the two centrality bins and compared to pp. The left panel shows the ratio for $R=0.2$ and $R=0.4$ and the right panel for $R=0.2$ and $R=0.6$. Within uncertainties, neither a centrality dependence nor a modification with respect to pp is observed. For $\sigma(0.2)/\sigma(0.6)$, the central values are slightly higher in central Pb--Pb collisions, but no significant enhancement is observed.

% RAA, comparison R=0.4 and R+0.6

% CSR

%%%%%%%%%%%%%%%%%%%%%%%%%%%%%%%%%%%%%%%%%%%%%%%%%%%%%%%%%%%%%%%%%%%%%%%%%%%%%%%%
\section{Conclusions}

In this paper, we presented transverse momentum spectra, nuclear modification factors, and cross-section ratios of track-based jets in Pb--Pb collisions at $\sqrt{s_\mathrm{NN}} = 5.02$ TeV that have been corrected by our novel Machine-Learning-based background correction approach. Thanks to the new background estimation method, jets with resolution parameter $R=0.6$ could be measured for the first time in Pb--Pb collisions at the LHC, jets with $R=0.4$ were measured down to 40 (30)~\GeVc\ for 0-10\% (30-50\%) most central collisions, unprecedented thus far in data on heavy-ion collisions.
Comparing the nuclear modification factors for different resolution parameters does not reveal a huge dependence of the jet modification on $R$, even for the largest $R=0.6$-jets.
This is also reflected in the jet cross-section ratios measured in Pb--Pb collisions: They do not show a strong deviation from the pp baseline result.

In a future analysis, it might be interesting to explore even larger jet resolution parameters. A comparison of the resolution parameter or radius dependence of jet nuclear modification to very recent Hybrid Model calculations~\cite{Hybrid2019} is promising. However, the maximum measurable jet $R$ in the ALICE detector is $R=0.9$ due to the limited track acceptance.

%%%%%%%%%%%%%%%%%%%%%%%%%%%%%%%%%%%%%%%%%%%%%%%%%%%%%%%%%%%%%%%%%%%%%%%%%%%%%%%%

\end{document}